\documentclass[aps,prl,twocolumn,showpacs,preprintnumbers,amsmath,amssymb,a4paper]{revtex4}
\usepackage{graphicx}% Include figure files
\usepackage{dcolumn}% Align table columns on decimal point
\usepackage{bm}% bold math

\begin{document}
\preprint{J. Y. Jo \textit{et al.}}
\title{Polarization Switching Dynamics Governed by Thermodynamic Nucleation Process \\in Ultrathin Ferroelectric Films}% Force line breaks with \\
\author{J. Y. Jo,$^{1}$ D. J. Kim,$^{1}$ Y. S. Kim,$^{1}$ S.-B. Choe,$^{1}$ T. K. Song,$^{2}$
 J.-G. Yoon,$^{3}$}
\author {T. W. Noh$^{1,}$}
\email{twnoh@snu.ac.kr} \affiliation{$^{1}$School of Physics and
Astronomy, Seoul National University, Seoul 151-747, Korea\\
$^{2}$School of Nano $\&$ Advanced Materials University, Changwon National University, Changwon, Gyeongnam 641-773, Korea\\
$^{3}$Department of Physics, University of Suwon, Suwon,
Gyeonggi-do 445-743, Korea}
\date{\today}% It is always \today, today,
\begin{abstract}
A long standing problem of domain switching process - how domains
nucleate - is examined in ultrathin ferroelectric films. We
demonstrate that the large depolarization fields in ultrathin
films could significantly lower the nucleation energy barrier
($U$*) to a level comparable to thermal energy ($k_{B}T$),
resulting in power-law like polarization decay behaviors. The
``Landauer's paradox": $U$* is thermally insurmountable is not a
critical issue in the polarization switching of ultrathin
ferroelectric films. We empirically find a universal relation
between the polarization decay behavior and $U$*/$k_{B}T$.
\end{abstract}

\maketitle

Recent advances in complex oxide thin film synthesis and in
first-principles calculations have intensified the basic research
on ferroelectricity at nanoscale dimension \cite{YSKim1,Junquera}.
The cooperative nature of ferroelectricity is expected to induce
different polarization states at nanoscale due to low-dimensional
boundary conditions and the strong interactions of polarization
with strain, charge, and other electromechanical parameters
\cite{Junquera,YSKim1,Fong,Choi,HNLee,Kornev,Lai}. Many workers
have reported intriguing physical phenomena occurring in ultrathin
ferroelectric (FE) films and other nanostructures, such as
intrinsic size effects \cite{Junquera,YSKim1,Fong},
strain-enhanced FE properties \cite{Choi}, unusual low-dimensional
phases \cite{HNLee}, and domain patterns \cite{Kornev,Lai}. On the
other hand, the mechanism and domain dynamics of FE switching in
ultrathin films have rarely been investigated in spite of their
scientific and technological
importance \cite{Tybell,Scott,So}. \\
\indent Historically, the mechanism of polarization switching
dynamics in ferroelectrics has been the subject of a great deal of
research. It is now believed that the polarization switching takes
place not by coherent switching, but by the nucleation and growth
of new domains \cite{Tybell,Scott,So}.  However, there still
remains an unsolved issue: How do the domains nucleate?  In the
late 1950s, Landauer emphasized that a thermodynamic nucleation
process cannot play a role in FE domain switching \cite{Landauer}.
For a nucleus with reversed polarization to be created by a
thermodynamic process, an energy barrier for nucleation $U$*
should be thermally overcome, as shown in Fig. 1(a). However,
Landauer's and later estimates showed that $U$* is practically
insurmountable by the thermal activation process: $U$* $>$
10$^{8}$ $k$$_{B}$$T$ at an electric field $E$ $\sim$ 1 kV/cm (a
typical value of the coercive field for bulk ferroelectrics)
\cite{Landauer} and $U$ $\sim$ 10$^{3}$ $k$$_{B}$$T$ at $E$ $\sim$
100 kV/cm (a typical value of the coercive field for most FE thin
films) \cite{So,Bratkovsky}, where $k$$_{B}$ is Plank's constant
and $T$ is temperature. According to these estimates, it is
difficult to understand the observed domain nucleation in
ferroelectrics. This problem has been known as the ``Landauer's
paradox". To overcome this difficulty, numerous workers have
assumed that the nuclei could be formed inhomogeneously due to
external effects, such as defects \cite{Aharoni}, long-range
interaction between nuclei \cite{Bratkovsky} and FE-electrode
coupling \cite{Gerra}. However, in this Letter, we will
demonstrate that $U$* could be thermally overcome in the ultrathin
FE film. The thermodynamic nucleation of reversed domains due to a
large depolarization field $E$$_{d}$ can govern the polarization
switching process, without inclusion of any
external effects, in ultrathin epitaxial BaTiO$_{3}$ films.\\
%==================[ Figure 1] =======================
\begin{figure}[floatfix]
\includegraphics[width=3.0in]{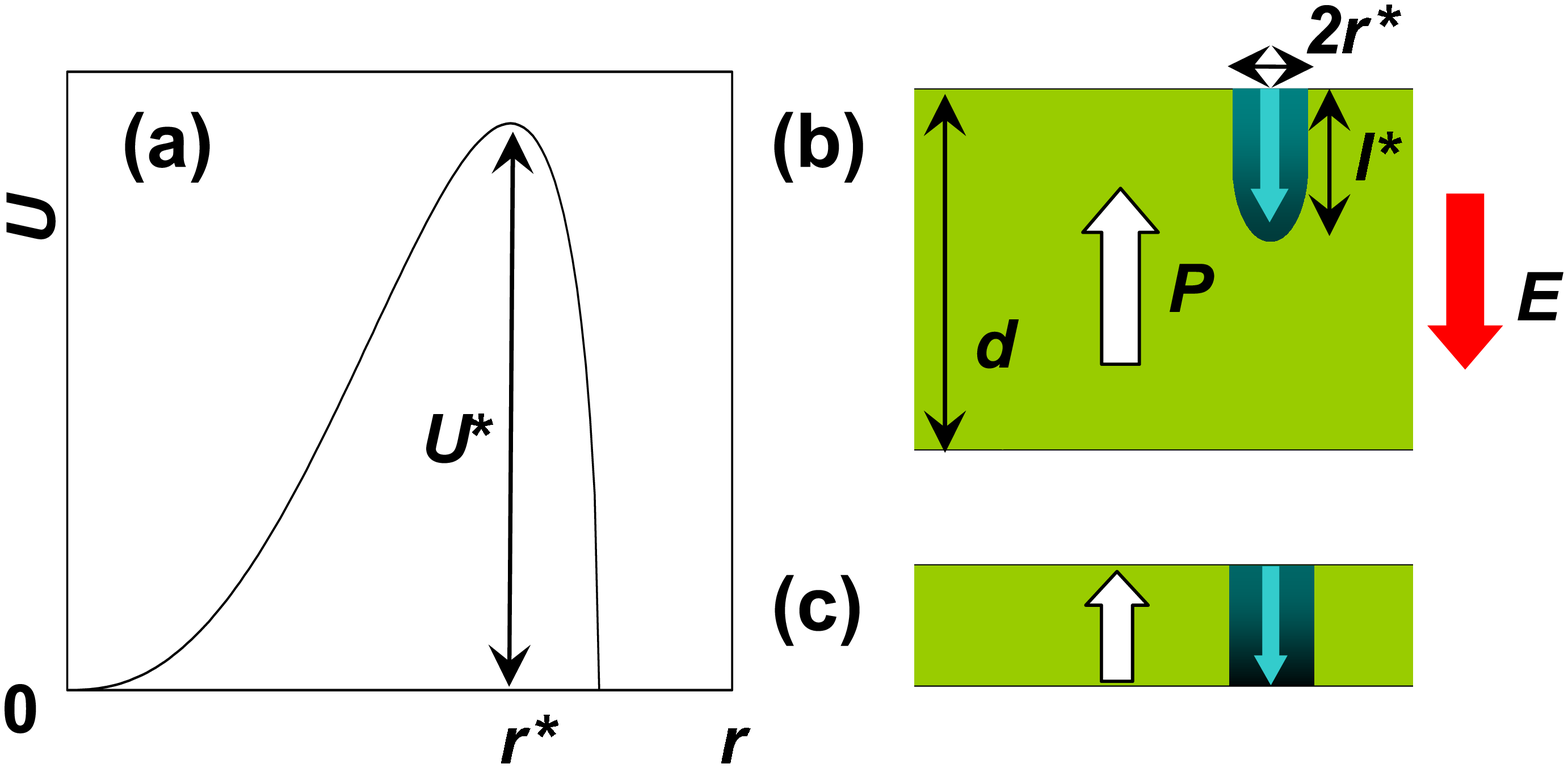}
\caption{\label{fig1 :epsart} (color online). (a) A schematic
diagram of electrostatic energy for the nucleus formation ($U$) as
a function of the nuclear radius ($r$). Schematic diagrams of (b)
the half-prolate spheroidal nucleus formation and (c) the
cylindrical nucleus formation, with reversed polarization.}
\label{Fig:1}
\end{figure}
%=====================================================
%==================[ Figure 2] =======================
\begin{figure}[floatfix]
\includegraphics[width=2.7in]{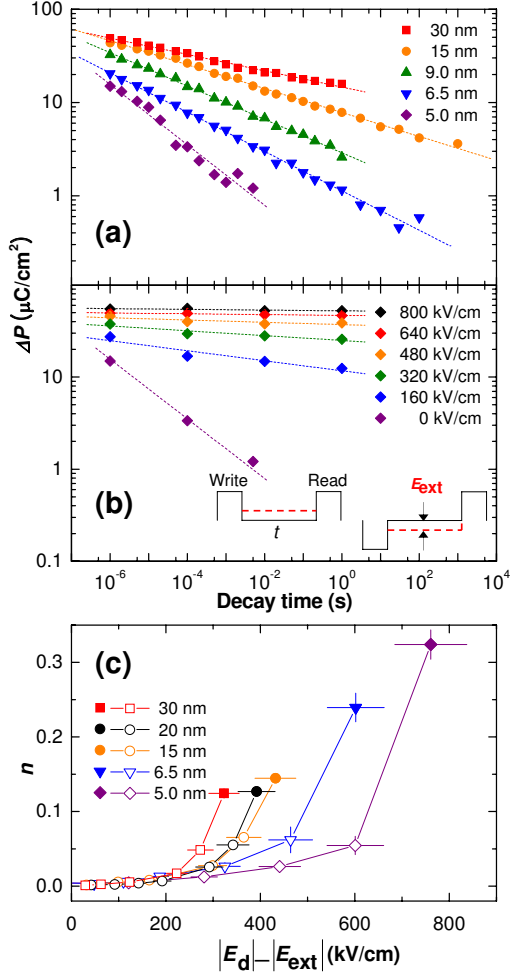}
\caption{\label{fig2 :epsart} (color online). (a) Time-dependent
net polarization changes, $\Delta$$P$($t$), without any external
field ($E_{ext}$). (b) $\Delta$$P$($t$) of the 5.0 nm thick
BaTiO$_{3}$ capacitor under numerous values of $E$$_{ext}$. The
left- and right-hand diagrams of the inset show electric pulse
trains used to measure non-switching and switching polarizations,
respectively. $\Delta$$P$($t$) corresponds to the difference
between these polarizations. (c) The relation between $n$ and the
total internal electric field (namely,
$|$$E$$_{d}$$|$-$|$$E$$_{ext}$$|$). The open and solid symbols
correspond to the data with and without $E$$_{ext}$,
respectively.}\label{Fig:2}
\end{figure}
%=====================================================
\indent Fully strained SrRuO$_{3}$/BaTiO$_{3}$/SrRuO$_{3}$
capacitors of high quality were fabricated using pulsed laser
deposition with reflection high-energy electron diffraction
\cite{YSKim1}. The thickness of BaTiO$_{3}$ ($d$$_{BTO}$) is
between 5.0 and 30 nm. We were able to directly measure $P$-$E$
hysteresis loops and polarization decay behaviors even for a 5.0
nm thick BaTiO$_{3}$ film \cite{YSKim1,DJKim,YSKim2}. We focused
on the polarization decay behaviors for different BaTiO$_{3}$ film
thickness and measurement conditions. As shown in Fig. 2(a), these
SrRuO$_{3}$/BaTiO$_{3}$/SrRuO$_{3}$ capacitors with a 10$\times$10
$\mu$m$^{2}$ area displayed an intriguing time-dependent
polarization decay behavior. We measured the time-dependence of
polarization decay using electric write/read pulses separated by a
given decay time ($t$) \cite{Scott,DJKim}: the right and the left
schematics in the inset of Fig. 2(b) show the electric pulses used
to measure the switching and non-switching polarizations,
respectively. Figure 2(a) shows that the polarization decay
follows a power-law behavior at room temperature \cite{DJKim}:
namely, $\Delta$$P$($t$) $\sim$ $t$ $^{- n}$. As the film becomes
thinner, the value of $n$ increases substantially. In particular,
for the 5.0 nm thick capacitor, $\Delta$$P$ decays below 1$\%$ of
its initial value within 10$^{-3}$ s. This rapid polarization
decay must be caused by domain backswitching upon removal of the
electric field and is
the subject of the discussions that follow.\\
\indent Within the capacitor geometry, the discontinuity of
polarization at the FE/electrode interfaces induces polarization
charges. The free carriers inside the electrodes compensate the
resulting charge deficit. However, the finite value of the
Thomas-Fermi screening length for the free carriers makes the
compensation incomplete, resulting in $E$$_{d}$ inside the FE
layer \cite{Mehta}. It should be noted that the generation of
$E$$_{d}$ is inevitable, which will limit the performance of most
FE capacitor-type devices.\\
\indent To obtain further information on the detailed effects of
$E$$_{d}$ on the polarization decay, we measured the
time-dependence of  $\Delta P$ with an external static field
$E$$_{ext}$, applied in the opposite direction to that of
$E$$_{d}$ during the electric pulse measurements, displayed
schematically by the red dashed lines in the inset of Fig. 2(b)
\cite{DJKim}. Since $E$$_{ext}$ will partially cancel out
$E$$_{d}$, it will slow down the polarization decay and
consequently reduce the values of $n$. As shown in Fig. 2(b), $n$
decreases significantly with an increase of $E$$_{ext}$ and should
finally become zero when $E$$_{ext}$ becomes nearly the same as
$E$$_{d}$. This $E$$_{ext}$ value at $n$ $\sim$ 0 provides an
experimental value for $E$$_{d}$ \cite{DJKim}. For the 5.0 nm
thick BaTiO$_{3}$ film, $E$$_{d}$ can be as large as 800 kV/cm.
Note that this value of $E$$_{d}$ is even larger than the coercive
field value, $i.e.$ about 300 kV/cm, which was measured from
direct $P$-$E$ hysteresis measurements \cite{YSKim1}.

Figure 2(c) shows the dependence of $n$ on the total internal
electric field, $i.e.$  $|$$E$$_{d}$$|$-$|$$E$$_{ext}$$|$, for the
BaTiO$_{3}$ capacitors. The abscissa-values of the solid symbols,
which were measured with $E$$_{ext}$ =0, correspond to the
experimentally estimated $E_{d}$ values. They clearly show that
$E$$_{d}$ becomes larger as $d$$_{BTO}$ becomes smaller. As the
$|$$E$$_{d}$$|$-$|$$E$$_{ext}$$|$ increases, the values of $n$
increases for all samples. However, we could not find any simple
universal relation between $n$ and
$|$$E$$_{d}$$|$-$|$$E$$_{ext}$$|$.

Further insights could be obtained by looking into the mechanism
of domain switching dynamics in an ultrathin FE film. Note that
decrease of $\Delta$$P$ means an increase in the number of FE
domains with a reversed polarization. It is generally believed
that the growth of the opposite domains occurs in three processes,
namely (a) formation of nuclei with opposite polarization, (b)
their forward growth, and (c) sidewise growth (called, the domain
wall motion). For epitaxial FE films thicker than 100 nm, the
domain wall motion has been reported to be very important
\cite{So}. However, for ultrathin FE films, domain wall motion
becomes extremely slow: for example, the speed of the domain wall
motion was estimated to be about 1 nm/s for a 29 nm thick
Pb(Zr,Ti)O$_{3}$ film \cite{Tybell}. In addition, the forward
growth of the nuclei is known to occur very fast ($i.e.$ a speed
of 1000 m/s) \cite{Scott}, so the very slow domain wall motion
could not play a major role in the polarization decay behavior
observed in Fig. 2(a). Therefore, we propose that, in the
ultrathin FE films, the rapid polarization decay should be caused
by nucleation-governed domain dynamics and that the large $E_{d}$
value could make the thermodynamic nucleation process possible.\\
\indent Figures 1(b) and 1(c) schematically display formation of
nuclei with half-prolate spheroidal and cylindrical shapes,
respectively, under an electric field. Here, $d$ is the FE film
thickness, and $r$* and $l$* represent critical radius and length
of the half-prolate spheroidal nucleus, respectively. Kay and Dunn
calculated the electrostatic energy $U$ for such nuclei formation
\cite{Kay}, and showed that there should exist the nucleation
energy barrier $U$*, as shown in Fig. 1(a). When $l$* $<$ $d$, the
spheroidal nuclei are likely to be formed, and $vice$ $versa$.
They also showed that $U$* is proportional to
$\sigma$$_{w}$$^{3}$$/$$E$$^{5/2}$ for the half-prolate nucleus
and $\sigma$$_{w}$$^{2}$$/$$E$ for the cylindrical nucleus
\cite{Kay}, where $\sigma$$_{w}$ is the domain wall energy. Using
$\sigma$$_{w}$ = 10 mJ/m$^{2}$ \cite{Merz}, the crossover from the
half-prolate spheroidal to cylindrical shapes will occur at $d$
$\sim$ 10 nm \cite{remark}. Note that the very large $E_{d}$
values of our ultrathin BaTiO$_{3}$ films will make $U$* small
enough to allow a thermally-activated nucleation process. The
experimental $E_{d}$ values of our BaTiO$_{3}$ films are between
300 and 800 kV/cm \cite{DJKim}. With these $E$$_{d}$ values, the
$U$*/$k$$_{B}$$T$ values are estimated to be between 4 and 20 at
room temperatures, which are easily accessible by thermal energy.

%==================[ Figure 3] =======================
\begin{figure}[floatfix]
\includegraphics[width=2.7in]{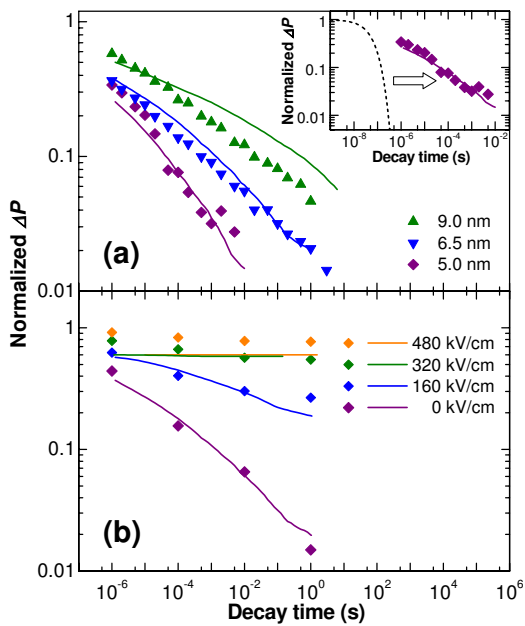}
\caption{\label{fig3 :epsart} (color online). (a) Normalized
$\Delta$$P$ from the experimental data (the solid symbols) and the
Monte-Carlo simulation (the solid lines). The black dotted line in
the inset shows the Monte-Carlo simulation results for the 5.0 nm
thick BaTiO$_{3}$ capacitor without lateral screening of the
interaction term between long-range domains in the Hamiltonian.
(b) Normalized $\Delta$$P$ of the 5.0 nm thick BaTiO$_{3}$
capacitor under numerous values of $E$$_{ext}$. The solid symbols
and lines correspond to the experimental data and the Monte-Carlo
simulation results, respectively.}\label{Fig:3}
\end{figure}
%=====================================================

\indent To explain the observed power-law decay behavior, we used
numerical simulations adopting the Monte-Carlo algorithm, which
takes into account the thermally-activated polarization reversal
process. We approximated the film to be composed of 128 $\times$
128 single domain cells lying on the $ab$-plane, whose size is 2.0
nm and height is the same as the film thickness, with a periodic
boundary condition. Each cell has a uniform $P$, either pointing
up or down along the $c$-axis. We used a dimensionless parameter
$x$ ($\equiv P/P_{s}$), and set the value of $x$ to be either +1
or -1 for the bi-stable states, where $P_{s}$ is spontaneous
polarization. During the switching process, the Hamiltonian
$H$$^{i}$ per volume $V^{i}$ of the $i$-th cell can be expressed
as a function of the $i$-th value of $x^{i}$;
%========= [Equation 1] ========================
\begin{eqnarray}
H^{i}=-K_{2}(x^{i})^{2}-\frac{\sigma_{w}d_{BTO}}{2V^{i}}x^{i}\sum_{n.n.}r_{n.n.}^{i}x_{n.n.}^{i} \nonumber\\
-E_{d}^{i} \cdot P_{s}x^{i}-E_{ext} \cdot P_{s}x^{i}.
\end{eqnarray}
%===============================================
The first self-energy term is quite similar to the leading $P^{2}$
term in the Landau-Devonshire equation, which provides an
approximation of the barrier height between the bi-stable points
in the FE potential \cite{Pertsev}. The higher order terms $P^{4}$
and $P^{6}$ were ignored here for the simplicity of the
calculations. The second term describes the short-range wall
interaction between the nearest neighboring domains, where
$r_{n.n.}$ and $x_{n.n.}$ are the adjacent circumference and unit
vector of nearest neighbor domain cells, respectively. It has a
maximum value when the 4 nearest neighbor domain cells surrounding
a nucleated cell have the opposite polarization. The third term
originates from the long-range dipolar interaction between cells.
The dipolar field $E_{d}^{i}$ is the sum of the dipolar electric
field caused by all the other cells. The last term describes the
work done by external source. A similar calculation method was
used in micro-magnetic systems \cite{SBChoe}. \\
\indent When two electrodes sandwich a FE film, the electrode
screens the value of $E_{d}^{i}$ produced by $P$$_{s}$. The
screening can occur not only in the perpendicular direction but
also in the lateral direction. A numerical estimate reveals that
the lateral screening length should be of the order of the FE film
thickness. For an ultrathin FE film, only the dipolar interaction
between the nearest neighbor domains across domain boundaries
could be important, so $E_{d}^{i }$ can be approximated to
$\hat{e}_{d}$ $\sum_{n.n.}$ $r_{n.n.}^{i}x_{n.n.}^{i}$, where
$\hat{e}_{d}$ is the line density of the dipolar field across the
cell boundary. Then, Eq. (1) can be rewritten as
%========= [Equation 2] ========================
\begin{eqnarray}
H^{i}=-K_{2}(x^{i})^{2}-\frac{\sigma_{w}^{(eff)}d_{BTO}}{2V^{i}}x^{i}\sum_{n.n.}r_{n.n.}^{i}x_{n.n.}^{i} \nonumber\\
-E_{ext} \cdot P_{s}x^{i},
\end{eqnarray}
%===============================================
where an effective wall energy density $\sigma_{w}^{(eff)}$ can be
defined as
$\sigma_{w}^{(eff)}=\sigma_{w}+\hat{e}_{d}P_{s}2V^{i}/d_{BTO}$.
Note that the analytic form of the screened dipolar interaction
energy term is identical with that of the domain wall energy. The
simulations were performed for room temperature domain dynamics
with $K$$_{2}$ = 3$\times$10$^{6}$ J/m$^{3}$,
$\sigma_{w}^{(eff)}$=
-1 mJ/m$^{2}$, and an attempt frequency of 10$^{9}$ Hz.\\
\indent The Monte-Carlo simulations were performed with and
without lateral screening of the interaction term between
long-range domains in the Hamiltonian. Without the lateral
screening of the interaction term, the simulations for the 5.0 nm
thick BaTiO$_{3}$ film predicted a very rapid exponential decay of
$\Delta$$P$, as shown by the black dotted line in the inset of
Fig. 3(a). When the screened domain interaction term was included
as in Eq. (2), the simulations showed that the polarization decay
would slow down significantly, exhibiting a power-law behavior, as
shown in the inset of Fig. 3(a). The calculated polarization decay
agrees with the experimental results quite well. Without changing
any other parameter values, we performed the same simulations for
other thicknesses of BaTiO$_{3}$ capacitors, and found reasonably
good agreement, as shown in Fig. 3(a). We also performed similar
simulations for the cases with non-zero values of $E$$_{ext}$, and
obtained good agreement with experimental data, as shown in Fig.
3(b). These results clearly demonstrate that domain dynamics must
be governed by thermally-activated nucleation processes in
ultrathin films \cite{remark}.\\
%==================[ Figure 4] =======================
\begin{figure}[floatfix]
\includegraphics[width=3.0in]{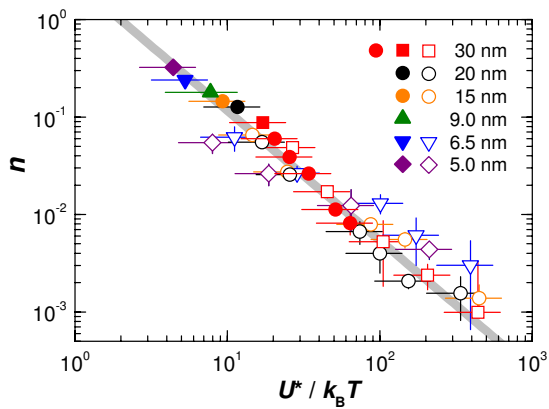}
\caption{\label{fig4 :epsart}(color online). A scaling relation
between $n$ and $U$*$/$$k$$_{B}$$T$. The open and solid symbols
come from the experimental $\Delta$$P$($t$) data with and without
$E$$_{ext}$, respectively.}\label{Fig:4}
\end{figure}
%=====================================================
\indent When the thermodynamic domain nucleation is the governing
process for polarization decay, $U$*/$k$$_{B}$$T$ should be the
most important physical term which determines the decay rate.
Following the methods in Ref. \cite{Kay}, we were able to convert
the values of ($|$$E$$_{d}$$|$ - $|$$E$$_{ext}$$|$) into $U$*
values and made a plot of $n$ vs. $U$*$/$$k$$_{B}$$T$. Figure 4
demonstrates that there might exist a universal relation between
$n$ and $U$*$/$$k$$_{B}$$T$, regardless of $d$$_{BTO}$ and
($|$$E$$_{d}$$|$ - $|$$E$$_{ext}$$|$). To confirm the
thermodynamic nucleation-governed polarization switching process,
we independently performed polarization decay measurements for the
30 nm thick BaTiO$_{3}$ capacitor at various temperatures. The
solid red circles in Fig. 4 correspond to the experimental data
for temperature-dependent polarization decay, which fall onto the
seemingly universal line between $n$ and $U$*$/$$k$$_{B}$$T$. This
indicates that there could be a simple relation between $n$ and
$U$*$/$$k$$_{B}$$T$, which is universal, irrespective of film
thickness, total internal electric field, and temperature.\\
\indent Our findings about the thermodynamic nucleation-governed
switching process and the empirical relation between $n$ and
$U$*/$k$$_{B}$$T$ have some important implications for the design
of nanoscale FE devices. One implication concerns the capacitor
geometry of FE devices where the polarization decay due to the
$E$$_{d}$ is predictable by the empirical relation we found. It
could impose a new fundamental thickness limit on such FE devices.
For example, ferroelectric random access memories require the
retention of a certain value of $\Delta$$P$ for 10 years
\cite{Scott}. For our BaTiO$_{3}$ capacitors, $U$* should be
higher than 40 $k$$_{B}$$T$ to retain $\Delta$$P$ at more than
50$\%$ of the initial value ($n$ $<$ 0.02). This value of $U$*
corresponds to a thickness of about 40 nm for our fully-strained
BaTiO$_{3}$ films, which is much thicker than the critical
thickness (2.4 nm) for the ferroelectricity of BaTiO$_{3}$
\cite{Junquera}. For ultrathin film capacitors, a condition of
$U$* $\geq$ $A$$\cdot$ $k$$_{B}$$T$, where $A$ is a constant of
the order of 10, has to be satisfied in order to have a stable
retention property. As the $E$$_{d}$ value increases (decrease in
$U$*) with the reduction of $d$$_{BTO}$, retention loss due to the
thermodynamic nucleation of reversed domains becomes much severe.
Note that this situation is analogous to that of the
superparamagnetic limit for magnetic memory devices: a long-range
ferromagnetic order vanishes, when anisotropy energy
becomes comparable to thermal fluctuation energy \cite{FM}.\\
\indent Our work has clearly shown the importance of a
thermodynamic domain nucleation process in the polarization decay
of ultrathin FE films. The polarization decay should impose a
fundamental thickness limit on capacitor-type devices, which is
much stricter than the critical thickness limit. Further
experimental and theoretical studies are clearly required to
clarify the origin of the empirically-determined relation between
$n$ and $U$*$/$$k$$_{B}$$T$ and its applicability to other
ultrathin FE systems.\\
\indent We thank M. W. Kim for discussion. We acknowledge
financial support by the Korean Ministry of Education through the
BK21 projects and the KOSEF through the CRI Program.


\begin{thebibliography}{30}
\bibitem{remark}
Our coercive field data agrees well with the prediction for the
crossover from the half-prolate spheroidal to cylindrical shapes
at $d$$_{BTO}$ $\sim$ 10 nm and also supports the picture of
polarization switching dynamics governed by thermodynamic
nucleation process in ultrathin ferroelectric films (to be
published).
\end{thebibliography}
\end{document}